\RequirePackage{silence}
\documentclass[reprint,aps,prx,floatfix,superscriptaddress]{revtex4-2}
\usepackage{amsmath,amssymb,amsthm}
\usepackage{graphicx}
\graphicspath{{figures/}{./}}
\usepackage{booktabs}
\usepackage{longtable}
\usepackage{xcolor}
\usepackage[hidelinks]{hyperref}
\usepackage{rotating}
\usepackage{seqsplit}
\usepackage{xr}

\counterwithout*{equation}{section}

\newtheoremstyle{thmnoparen}{6pt}{6pt}{\itshape}{}{\bfseries}{.}{ }{\thmname{#1}\thmnumber{ #2}\thmnote{ #3}}
\theoremstyle{thmnoparen}

\DeclareFontShape{OT1}{cmr}{bx}{sc}{<->ssub*cmr/bx/n}{}
\newcommand{\Lean}{\textsc{Lean}}
\newcommand{\code}[1]{\texttt{\seqsplit{#1}}}
\providecommand{\coloneqq}{\mathrel{\mathop:}=}

\begin{document}

\title{Quantum error-correcting code parameters, checkable by a certificate of provable size}

\author{Shuoming An}
\email{anshuoming@suat-sz.edu.cn}
\affiliation{Faculty of Computility Microelectronics, Shenzhen University of Advanced Technology, Shenzhen, Guangdong 518107, China}
\affiliation{Guangdong Provincial Key Laboratory of Computility Microelectronics, Shenzhen, Guangdong 518107, China}
\affiliation{Pengxin Quantum Technology (Shenzhen) Co., Ltd., Shenzhen 518107, China}
\author{Fusheng Yang}
\email{yangfsh7@mail2.sysu.edu.cn}
\affiliation{Institute of Quantum Computing and Software, School of Computer Science and Engineering, Sun Yat-sen University, Guangzhou 510006, China}

\begin{abstract}
Quantum error correction protects quantum information by encoding it in a code described by three numbers: how many physical qubits it uses, how many logical qubits it protects, and the smallest error it cannot detect. The last is an optimum over exponentially many candidates, so it is read off a solver rather than checked. Here we show that code parameters can be certified by a short certificate, a list or a pairing or a symbolic instance, whose correctness the kernel of a proof assistant decides by computation, and whose size is itself a theorem rather than an estimate. Two published code families have distances settled here: the bivariate bicycle code's $[[144,12,12]]$, certified, and the affine-permutation codes printed at $[[1152,580,\le 12]]$ and $[[2304,1156,\le 14]]$, settled at twelve and fourteen. Eleven code families and thirty-nine parameter sets follow, with nothing beyond three standard axioms trusted and the pipeline's specification step preset to twenty columns.
\end{abstract}

\maketitle

\section{Introduction}
\label{sec:intro}

Quantum error correction keeps a quantum computer reliable by correcting errors instead of
dodging them. The information is encoded across many physical qubits, and the checks that
detect errors are read often enough that errors are caught before they pile up~\cite{shor1995scheme,steane1996error,steane1996multiple,calderbank1996good}. The best-studied family is the surface code, which sits on a two-dimensional lattice, checks only nearby qubits, and grew out of the toric code~\cite{kitaev2003fault,fowler2012surface}; the field as a whole is surveyed in Ref.~\cite{terhal2015quantum}. These codes have started to pay off in hardware. A distance-three surface code corrected errors repeatedly~\cite{krinner2022realizing}, a logical qubit outlived its best physical qubit~\cite{sivak2023realtime}, a larger code suppressed errors more strongly~\cite{google2025quantum}, and reconfigurable atom arrays ran a logical processor~\cite{bluvstein2024logical}.

How good a code is comes down to its distance, the weight of the lightest logical operator.
That weight is the fewest qubits an operation on the encoded information can touch and still
slip past the checks. The distance is an optimum over an exponentially large set, so in
practice it is read off an external solver. A solver's answer is reproducible without being
checkable: rerunning it tests the program rather than the claim. The difficulty is intrinsic,
since computing a stabilizer code's distance exactly, or even approximately, is
NP-hard~\cite{kapshikar2023hardness}. Hardness rules out a fast algorithm, not a short
certificate, so the routes that exist are the ones to build on, not the ones to replace.

Throughout, $[[n,k,d]]$ records a code's physical qubits, its logical qubits and its distance.
Two codes set the target. The bivariate bicycle code $[[144,12,12]]$, a
Calderbank--Shor--Steane code (CSS for short~\cite{steane1996error,steane1996multiple,calderbank1996good}), opened the era of low-overhead quantum memory by storing data with far fewer physical qubits than earlier designs needed~\cite{bravyi2024highthreshold}. Its checks come from two polynomials in two commuting variables, and it belongs to the quantum low-density parity-check family, which does what the surface code cannot. As the block length grows, the rate stays above a constant and the distance grows with it~\cite{tillich2014qldpc,breuckmann2021quantum,panteleev2021degenerate}. Its reported $d = 12$ came out of a mixed-integer program, an optimization in which some variables are restricted to whole numbers, in the original presentation~\cite{bravyi2024highthreshold}, and no certificate of the number was published with it. Codes of this family are searched as a matter of course: one guided search screened $2\times10^{5}$ candidates in about $140$ hours~\cite{evolvebb2026}, and a distance settled this way inherits the weakness that a reader cannot check it.

The second code is larger, and past the reach of a distance settled by enumeration. QuEra
Computing, Harvard and MIT built it in 2026 from affine permutations $x \mapsto ax + b$, on
reconfigurable neutral-atom arrays whose atoms can be moved while coherence lasts~\cite{quera2026apm}. In each check
block the qubits are relabelled by a fixed multiplier and offset. The largest instance has rates
above one half. Their $[[1152,580,\le 12]]$ instance appears in the appendix of that work, which
gives the affine data that generates it and the check matrices it generates. Its distance
appears there as a best-found upper bound from a randomized search, not an exact
value~\cite{quera2026apm}; this work carries the instance over and settles it at exactly twelve.
The construction is published along with the parameters, which is what lets a certificate be
built for this instance. An exhaustive check would have to walk a weight-limited candidate list
of $10^{16}$ entries at $n = 144$ and about $10^{26}$ at $n = 1152$. That is why the numbers
in the field's tables are printed as bounds rather than as exact values. Settling distances by
search has not kept pace with the codes, and the gap widens as they grow. Closing it needs a
change of unit rather than a faster search, since what has to scale is the certificate, not the
search that produced it.

Three routes to a distance are in use, and each drops one of the three things a certificate
would need: scale, generality, and a size theorem. The first reduces the question to
satisfiability and discharges it by bit-blasting, the translation of the question into Boolean
logic, and replays the solver's proof inside the \Lean\ proof assistant, the small trusted
core that checks each step. It reaches ninety qubits there, and 144 qubits only outside the
kernel, where its authors expect the replay, rather than the search, to be the next
bottleneck~\cite{ehatamm2026leanqec}. Scale is what it drops, since the replay is what sets the
ceiling.

The second route is written for one family at a time. The homological analysis, the algebra of
the chain complexes underneath a code, computes its distance in closed form, and reaches
$d = 12$ with a kernel-only
proof~\cite{jain2025qeclean}. The analytic routes outside the kernel share the shape, an exact
distance law for cyclic covers that carries one certified floor up an entire
tower~\cite{young2026distance}. Generality is what it drops, since each new family costs a new
proof.

The third route searches. It may enumerate completely, reduced by a symmetry whose action on the
checks has itself been verified and closed by an explicit witness~\cite{okada2026girth8}. Or it
may run on a platform that certifies codes as it finds them, the extended version of that
work adding an independent \Lean\ verification layer~\cite{he2026codesign}. Either way it hands back a computation, or an object no larger
than the instance. A size theorem is what it drops. A separate line verifies the programs that
operate on a code, with the code's parameters taken as given~\cite{huang2025veriqec}.

None of the three hands a reader an object short enough to inspect and strong enough to settle
the distance. What has not been put together is the pair: a certificate whose size is a theorem,
checked by the kernel, and produced per instance rather than per family. Whether checking a
code's parameters can be made uniform across codes, with a certificate of proved size, is open.
That per-instance pair is what this work supplies.

Here we show that it can, by changing the unit of verification: not the code and not the
family, but the \emph{certificate}. A certificate is a short object whose correctness the kernel
decides by computation, and it returns what each of the three routes drops. It makes checking
cheaper than searching: the kernel verifies a short object where the search would do the work. Two constructions carry it, one for each bound. The first construction carries the upper
bound. Row-space membership is the operation that would otherwise keep this route out of the
kernel at scale, and a single inner product against a second vector replaces it. To certify that
a vector $v$ lies outside the row space of the check matrix $H$, the set of sums of its rows,
exhibit $w$ with $Hw = 0$ and $w \cdot v = 1$. That is one matrix-vector product and one inner
product, with no row reduction. On the toric instance $[[18,2,3]]$, a surface code closed into a
torus, the same assertion's check falls from $42$\,s to $9$\,s. Where the elimination itself
is replaced rather than the certificate form, the change is starker: a computation that does
not converge at $200$\,s and $41$\,GB finishes in $8$\,s.
The second construction carries the lower bound. For product families it is proved once, symbolically
in the family parameter, by a cleaning argument, in which a low-weight logical operator is
pushed into a restricted shape by adding local checks. Whole families are then corollaries that
generate no candidate set of vectors, and the enumeration the certificate replaces has length
exactly $\sum_{j<d}\binom{n}{j}$, polynomial in $n$ at fixed $d$ rather than $2^n$.

The case for the change of unit is not confined to code parameters. Quantum error
correction is now an engineering program. A threshold is estimated by simulating a
family of codes as it grows, and the distance sets how fast the logical error rate
falls. The distance is one of the code parameters a certificate settles, so making it
checkable changes what such an estimate rests on. Code design is increasingly a machine
search, which produces candidates faster than it can verify them.
And the shape of the argument travels: wherever a scientific computation ends in an
external solver, a certificate the kernel decides is the part a reader can carry away.
The route runs inside the kernel, with no solver in the trusted base and no compiler trusted
where the checker decides, and no audited statement depends on anything beyond the three
standard axioms of \Lean's own logic. The dataset reaches eleven code families and
thirty-nine parameter sets, the widest at $n = 1872$ with $d = 7$. Uniformity has a width: the
pipeline's specification step, which writes a code's \Lean\ definition, carries a pre-set width
of twenty columns, and the widest instance the enumeration route covers, $n = 36$, had to be
written by hand.

The evidence is of three kinds, kept apart throughout: kernel
assertions, an independent recomputation of a sample of the parameter
sets, and wall-clock measurements on one machine. Sections~\ref{sec:setup}--\ref{sec:size} give the
representation, the reduction and the certificate; Section~\ref{sec:evidence} weighs them
against the evidence forms in use; Sections~\ref{sec:structured}--\ref{sec:discovery} the
families, the instances and the two pipelines; Section~\ref{sec:audit} what the kernel
trusts, and Section~\ref{sec:limits} the boundary of the claims.

\section{Results}

\begin{figure*}[t]
\includegraphics[width=0.95\textwidth]{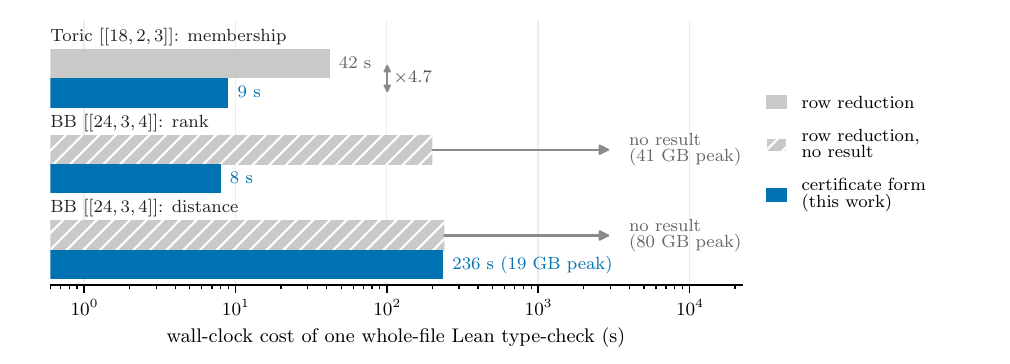}
\caption{The verification method affects run time. Each row compares two ways of
checking the same claim: the first row changes the certificate form, and the other two
hold it fixed and change the elimination that computes it. Bars show the time to
type-check the full Lean module for that claim: gray for the earlier method and blue
for the method used here; the time axis is logarithmic, so bar length is not
proportional to time. The first row tests whether a weight-3 logical operator of the
toric code lies outside its stabilizer row space; the other rows show rank and
distance-lower-bound checks for the $[[24,3,4]]$ code.
Hatching marks a run stopped without a result; its label gives the stopping time and
peak memory, so the time is a lower bound. All runs used one machine and a fixed memory
limit; the distance check completed within it.}
\label{fig:cost}
\end{figure*}
\subsection{Setting and representation}
\label{sec:setup}

\paragraph{Codes as GF(2) row spaces.}
A stabilizer code is defined by a set of mutually commuting Pauli operators.
In a CSS code, the stabilizer generators can be chosen as two sets: $X$-type checks
containing only $I$ and $X$, and $Z$-type checks containing only $I$ and $Z$.
These sets are represented by two binary parity-check matrices, $H_X$ and $H_Z$,
respectively.
Every question we ask is then
a question about a subspace of $\mathbb{F}_2^{n}$, written GF(2) below, the two-element
field whose vectors are bit strings added modulo two. We therefore work directly with
$\mathrm{Vec}(n) \coloneqq \mathrm{Fin}(n) \to \mathbb{F}_2$, so that vectors are
functions on a finite index set rather than bit strings, and a code is the row space of a
list of such vectors. With this representation, membership in a row space can be
decided by a finite computation that \Lean's proof kernel can verify.

\paragraph{Interface.}
We prove in \Lean\ that our definitions of row spaces, Pauli commutation and Pauli
weight agree with the corresponding definitions in existing Lean libraries.
We also prove that any state of the logical register has von Neumann entropy at most
$k$ bits, where $k$ is the certified number of logical qubits.
The supplement lists the corresponding theorems, all checked by \Lean's proof kernel.
When different libraries use the same name for different objects, the interface
documentation specifies which library's definition is used.
The GF(2) definitions, Pauli-operator representation and shared code instances are
released separately in the QECCertificates package.
This project uses a fixed version of that package to make the proofs reproducible.
The Data Availability statement identifies the release.

\paragraph{What is checked.}
For a CSS code specified by the parity-check matrices $H_X$ and $H_Z$, we aim to verify
its number of logical qubits $k$ and its distance $d$. With $\mathrm{wt}$ the Hamming
weight, the number of positions at which a vector is nonzero, $\operatorname{rowspace}$ the
span of the rows, $\ker$ the null space, the set of
vectors a matrix sends to zero, and the width of a matrix its
number of columns, the quantities reported are

\begin{equation}
  \begin{aligned}
    k &= n - \operatorname{rank}H_X - \operatorname{rank}H_Z , \\
    d_X &= \min\bigl\{\mathrm{wt}(v) \;:\; v \in \ker H_X, \\
       &\qquad\qquad\quad\ v \notin \operatorname{rowspace}H_Z \bigr\},
  \end{aligned}
  \label{eq:definitions}
\end{equation}
with $d_Z$ the same expression under the exchange $H_X \leftrightarrow H_Z$, so that
$d = \min(d_X,d_Z)$. A classical code is the case of an empty $H_Z$, and its distance is
the first expression alone, since the second would be measuring the complement of the
code. The length-three repetition code is the smallest instance: its two check rows $110$
and $011$ have rank two, so $k = 3-2 = 1$, and the only nonzero vector they annihilate is
$111$, of weight three, so $d = 3$.
In our notation, the subscript in $d_X$ refers to the check matrix $H_X$:
$d_X$ is the minimum weight of a nontrivial $Z$-type logical operator.
Similarly, $d_Z$ is the minimum weight of a nontrivial $X$-type logical operator.
These labels are exchanged relative to the usual error-type convention;
Section~\ref{sec:instances} discusses the instances for which the two side distances differ.

For instances checked by enumeration, we obtain $k$ from Eq.~\eqref{eq:definitions}
after computing the ranks of $H_X$ and $H_Z$ by row reduction.
For instances built as hypergraph products, codes grown from two smaller seed
matrices, the dimension formula in Eq.~\eqref{eq:kunneth} gives $k$ from the two seeds,
without row reduction on the full check matrices.
To prove that the distance is $d=d_0$, we establish both $d\le d_0$ and $d\ge d_0$.
An upper bound $d\le d_0$ is proved by exhibiting a nontrivial logical operator of weight $d_0$.
A lower bound $d\ge d_0$ requires proving that no nontrivial logical operator has weight less than $d_0$.
The latter requires excluding all lighter candidates, either by exhaustive enumeration
or by a proof using the code's structure.

Both distance bounds for the thirty-nine parameter sets in our dataset are verified by
\Lean's proof kernel.
For the bivariate bicycle code $[[144,12,12]]$, the lower-bound proof comes from an
independent \Lean\ formalization; we prove that it applies to the same check matrices used here.
For the additional affine-permutation code $[[1152,580,12]]$, \Lean\ verifies the upper bound,
but the reported lower bound comes from an exhaustive computation outside \Lean. For the
larger affine-permutation code $[[2304,1156,14]]$, the upper bound is the best-found value
reported by the original work, and the lower bound comes from an exhaustive computation
outside \Lean.

\subsection{Reduction inside the kernel}
\label{sec:reduction}

To verify a claimed code parameter, we check explicit mathematical data, such as a
candidate logical operator together with evidence that it is nontrivial.
We call the data supplied for such a check a certificate.
For the enumeration route, we implement the checks in \Lean\ and prove that their
results imply the claimed parameter bounds.
Two basic operations are computing a matrix rank and deciding whether a vector lies
in a row space.
Both operations use row reduction, with the results verified by \Lean's proof kernel;
Figure~\ref{fig:cost} shows what these checks cost and how the way each is written sets
the cost.

\paragraph{Row reduction.}
During row reduction, we keep selected rows and associate a pivot column with each one.
In that column, the row has a $1$.
We call the rows mutually reduced when each row is $1$ in its own pivot column and $0$
in every other row's pivot columns.
For example, start with rows $110$ and $011$, using the first and second columns as
their pivots.
The first row is $1$ in the second pivot column, so add the second row to it:
$110+011=101$.
The resulting rows, $101$ and $011$, satisfy the condition.
Swapping the order in which the pivot rows are listed does not change whether this
condition holds.
The correctness proof therefore does not rely on a particular ordering of rows.
One theorem proves that row reduction preserves the span of the input rows, so the code's
row space is unchanged.
A second proves that the output rows satisfy the mutual-reduction condition above.

For wide matrices, we also use a second, append-only routine.
Unlike the first routine, it does not go back to update older pivot rows when it adds a
new one.
Instead, each new pivot row is reduced only in the columns that were chosen as pivots
earlier; all earlier rows stay as they are.
This second routine returns the same rank as the first.
Its rows need not be mutually reduced: for testing row-space membership, it is enough
that the pivot columns progress from left to right down the list.
To test a vector, we subtract the pivot rows needed to clear its pivot-column entries.
If the remainder is zero, the vector is in the row space; otherwise it is not.

Let $\operatorname{echelon}(M)$ denote the pivot rows returned when reducing $M$.
For a vector $v$, let $\mathrm{red}_M(v)$ be the remainder after those rows are used to
clear its pivot-column entries.
The two facts we use about these quantities are

\begin{equation}
  \begin{aligned}
    \operatorname{rank}M &= \bigl|\operatorname{echelon}(M)\bigr| , \\
    v \in \operatorname{rowspace}M &\iff \mathrm{red}_M(v) = 0 ,
  \end{aligned}
  \label{eq:reduction}
\end{equation}
The supplement proves both equations.
For the rank equation, row reduction preserves the row space and its output pivot rows
are independent, so their number is the rank.
For the membership equation, the pivot columns determine which rows to subtract, and
the remainder is zero exactly when $v$ lies in the row space.
The same pivot data also give a basis for the kernel of a check matrix.
Moreover, two matrices with the same row space reduce to the same canonical form.
We can therefore check whether two matrices define the same code by comparing these
forms, without testing all $2^n$ vectors; a certificate can identify a code by its form
instead of listing its matrix.

\paragraph{Scale limit.}
Executability has two costs. The first comes from how a matrix is represented. In Lean,
a matrix can be a function: give it a row position and a column position, and it returns
the entry at that position. The same matrix can instead be written as an explicit list
of its rows. For a matrix with $18$ columns, the same assertion takes more than five
minutes to check with the function representation, but about $95$ seconds with the row-list
representation.

We therefore state these assertions using explicit row lists. When a matrix starts in
the function representation, we also prove that it equals the corresponding row list;
Lean's kernel checks that equality. This lets us use the faster representation without
changing which matrix or mathematical claim is being checked.

The second cost comes from the elimination algorithm. In full row reduction, each time
a new pivot is found, the algorithm goes back and updates all the earlier pivot rows to
clear the new pivot column. As the list of pivot rows grows, each new step revisits more
old rows, so these repeated updates and copies quickly become expensive. On a matrix
with $24$ columns, full row reduction does not finish within practical limits, whereas
the append-only routine finishes in seconds to minutes. It avoids the repeated work by
leaving earlier rows unchanged and adding only the newly reduced row.

The same bottleneck can arise in other proofs of distance lower bounds that use
candidate enumeration. To prove that a code has distance at least $d$, one must rule out
every nontrivial logical operator of weight less than $d$. In certificate terms, this is
an emptiness check: none of the low-weight candidates is a logical operator.

\paragraph{A measured example: the $[[24,3,4]]$ code.}
The measured code has parameters $[[24,3,4]]$. It is obtained from the bivariate
bicycle code $[[18,4,4]]$ by promoting one logical operator, of weight four, to a
stabilizer; the promotion gives each of the six pairs of the operator's support qubits
an ancilla of its own, and those ancillas take the length from eighteen to twenty-four.
With the append-only routine, the rank check finishes in $8$\,s. The distance check
takes about four minutes and reaches a peak memory use of about $19$\,GB. Its parameters
also appear among the kernel-audited instances of Section~\ref{sec:instances}.

Figure~\ref{fig:cost} shows these measured times. It compares ways of establishing the
same mathematical claims: changing the certificate form for the toric-code example, and
changing the elimination algorithm for the rank and distance checks of the gauged code.

\begin{table*}[t]
\caption{Certificate size. For enumeration rows, the count is the number of vectors of
weight below $d$; $2^n$ gives the size of the full vector space for comparison. The three
largest rows use a symbolic distance proof in the cycle length $m$, so no candidate set is
generated; the proof uses two factor matrices of size $O(m^2)$. For the gauged
$[[24,3,4]]$ code, append-only row reduction makes the complete certificate check
tractable. Here HGP abbreviates the hypergraph product code, BB the bivariate bicycle code,
and a classical code is written with single brackets, $[n,k,d]$.
\label{tab:certcost}}
\begin{ruledtabular}
\begin{tabular}{lrrrrl}
Instance & $n$ & $d$ & $\sum_{j<d}\binom{n}{j}$ & $2^{n}$ & Route \\
\hline
Hamming $[15,11,3]$ & 15 & 3 & 121 & $3.3\times10^{4}$ & enumeration \\
BB $[[18,4,4]]$ & 18 & 4 & 988 & $2.6\times10^{5}$ & enumeration \\
Toric $[[18,2,3]]$ & 18 & 3 & 172 & $2.6\times10^{5}$ & enumeration \\
BB $[[24,3,4]]$, gauged & 24 & 4 & 2{,}325 & $1.7\times10^{7}$ & enumeration, append-only \\
HGP(paw, diamond) & 36 & 3 & 667 & $6.9\times10^{10}$ & enumeration \\
BB $[[144,12,12]]$ & 144 & 12 & $1.0\times10^{16}$ & $2.2\times10^{43}$ & transported analytic lower bound \\
HGP toric $m=9$ & 162 & 9 & $1.0\times10^{13}$ & $5.8\times10^{48}$ & structured, symbolic in $m$ \\
HGP toric $m=12$ & 288 & 12 & $2.4\times10^{19}$ & $5.0\times10^{86}$ & structured, symbolic in $m$ \\
HGP toric $m=16$ & 512 & 16 & $2.8\times10^{28}$ & $1.3\times10^{154}$ & structured, symbolic in $m$ \\
\end{tabular}
\end{ruledtabular}
\end{table*}

\subsection{Certificate forms}
\label{sec:certs}

Three certificates cover every instance result the enumeration route produces, and a
fourth form covers the pipeline's seed matrices, the two small matrices a product code is
grown from. Each is
short relative to the space it replaces, and the kernel checks each by computation; the
first three are complete for the property they certify, meaning that whenever the property
holds a certificate of that form exists. Their size is
a theorem about the algorithm, stated in Section~\ref{sec:size} and reported in
Table~\ref{tab:certcost}.

\paragraph{Rank certificate.}
The rank of a parity-check matrix is certified by an elimination the kernel runs itself:
the matrix goes in, a reduced form comes out, and its length is the rank. That the result
is reduced is a theorem about that elimination. Above the width the elimination reaches,
the rank comes from a submitted echelon form instead. This settles $k$, since
$k = n - \operatorname{rank}H_X - \operatorname{rank}H_Z$.

\paragraph{Dual witness.}
The upper bound on $d$ exhibits a logical operator of weight $\ell$, and deciding that a candidate is one is a membership question: it is in the kernel of the check matrix, and it
is not in the row space. Deciding the second half by row reduction is what forces this step
out of the proof assistant's kernel at scale, and the certificate replaces it with a
pairing. To prove that a
weight-$\ell$ vector \emph{is} a logical operator it suffices to prove it is in the
kernel of the check matrix and to exhibit a dual witness, a vector orthogonal to every row of the parity
check and pairing to $1$ with it:

\begin{equation}
  \begin{aligned}
    v \in \ker H , \quad Hw = 0 , \quad w \cdot v = 1
    \;\Longrightarrow\;
    v \notin \operatorname{rowspace}H ,
  \end{aligned}
  \label{eq:dualwitness}
\end{equation}
the reason being one step of linear algebra: every vector of the row space is
orthogonal to every vector of the kernel, and $w$ is a kernel vector, so $w \cdot v = 1$
places $v$ outside the row space. The smallest instance is two bits long: for $H$ the
single row $11$ the row space is $\{00,11\}$, and the witness $w=11$ excludes the
candidate $v=10$ through $Hw = 1+1 = 0$ and $w \cdot v = 1$, two additions in all.
The certificate's cost is the left-hand side of Eq.~\eqref{eq:dualwitness}: one
matrix-vector product and one
pairing rather than an echelon form. Its effect on wall-clock type checking is
measured in Section~\ref{sec:evidence}.

\paragraph{Emptiness certificate for the lower bound.}
The lower bound needs to exclude every undetectable logical operator of weight below
$d$, which is the exponential step. Where enumeration is chosen, the certificate is that
the candidate set is empty. The candidates are drawn by weight rather than from the whole
space: the list they are drawn from holds the vectors of weight at most $d-1$, of which there are
$\sum_{j<d}\binom{n}{j}$. The
reliability of this reduction is not assumed but proved, by a covering theorem
showing that every vector of weight at most a given bound appears in the generated list. The
generator itself is written as a weight-recursive enumeration, and the counting
theorem is stated about it, so the certificate size is a theorem.

\paragraph{One certificate, two answers.}
A fourth form appears in the pipeline of Section~\ref{sec:discovery}, where what must be
certified is a seed matrix rather than a code. A spanning certificate for a seed
matrix $M$ is a triple of small matrices whose product identities, over $\mathbb{F}_2$,
show that every kernel vector of $M$ is a combination of the columns of the first and that
the coefficients of that combination are uniquely determined. That replaces a null-space
search over $\sum_{j<d}\binom{n}{j}$ vectors by one over at most $2^{13}$ coefficient
vectors, and the same three identities make the containment an equality, so a seed's
dimension and its rank are read off the certificate rather than by a row reduction at the
seed's width. Both readings are kernel theorems, checked on the seed in one kernel
evaluation. The supplement gives the matrices' shapes, the identities, and the three steps
from them to the kernel.

\subsection{Certificate size}
\label{sec:size}

A certificate is only an improvement over a search if it is smaller, so we prove the
size. The weight-limited list for a length-$n$ code of distance $d$, the vectors
$v$ with $\mathrm{wt}(v)<d$ where $\mathrm{wt}$ is the Hamming weight, has

\begin{equation}
  \bigl|\{v : \mathrm{wt}(v) < d\}\bigr| = \sum_{j<d}\binom{n}{j}.
  \label{eq:candcount}
\end{equation}

Each term of that sum is at most a power of the length: since
$\binom{n}{j} \le n^{j} \le n^{\,d-1}$ for every $j < d$ and $n \ge 1$,

\begin{equation}
  \sum_{j<d}\binom{n}{j} \;\le\; \sum_{j<d} n^{\,j} \;\le\; d\,n^{\,d-1},
  \label{eq:polynomial}
\end{equation}
so for fixed $d$ the count is a polynomial of degree $d-1$ in $n$ rather than $2^{n}$. At
$d=3$ the sum is $1 + n + n(n-1)/2$, which at the length-fifteen Hamming code is $121$. The
candidate set the algorithm enumerates is that set filtered by the undetectable
predicate, so its size is at most this count; the enumeration itself has length
exactly the count. Table~\ref{tab:certcost} reports both columns for the
instances.

The bound is tight: the weight-limited list has that length for every code, and the
candidate set of the code with no parity checks at all, where every nonzero vector is
undetectable, keeps the entire list bar the zero vector, of length exactly
$\sum_{j<d}\binom{n}{j}-1$, so no bound
uniform over codes in both $n$ and $d$ can be smaller. Nor can the enumeration be pruned
away in general. The size
theorem therefore reads in both directions, the list length $\sum_{j<d}\binom{n}{j}$ as an
upper bound for every code and $\sum_{j<d}\binom{n}{j}-1$ attained by one. For a structured code the enumeration is avoided rather than
bounded, since no candidate set is generated at all.

One further cost is invisible in the
asymptotics: converting the weight-limited list into a set is quadratic in the number of
candidates, and at the largest enumerable instance that cost exceeded the decision
procedure it was preparing input for, a measurement the supplement carries. In a
certificate-based method the representation of the certificate is part of the algorithm,
which is why the interface offers both a set-typed and a list-typed entry point.

\begin{figure*}[t]
\includegraphics[width=12.1cm]{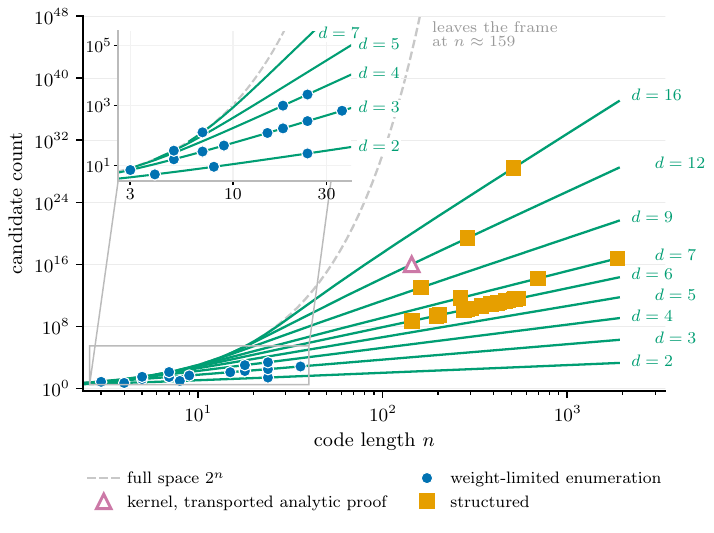}
\caption{The certificate stays polynomial while the space searched is exponential. Solid curves: the weight-limited
certificate $\sum_{j<d}\binom{n}{j}$ of Eq.~\eqref{eq:candcount} at fixed distance $d$, a
polynomial of degree $d-1$ in $n$; dashed: the full space $2^{n}$. The horizontal axis
is the code length $n$, the vertical a count of candidates, both on logarithmic scales.
Markers are colored and shaped by the route that certifies them; the structured route of
Section~\ref{sec:structured} and the graph-seed route of Section~\ref{sec:discovery}
share one color and one legend entry. A marker's height is a count of candidates, and for the
structured markers and the transported one, whose bound is imported from an independent
formalization, that count is the enumeration the certificate replaces rather than the size of
the certificate itself. The inset redraws the bottom-left corner at its own scale, where
each marker certified by enumeration can be read against the curve of its own $d$. The
separation opens as soon as $d$ is fixed, exceeding 500 orders of
magnitude at the largest instance, $n=1872$; that count is computed rather than drawn,
since the full-space curve
leaves the frame at $n\approx159$, as the figure annotates. }
\label{fig:certsize}
\end{figure*}

Fig.~\ref{fig:certsize} plots the same two columns for fixed $d$ as a function of $n$,
so that the separation between the two growth rates is visible rather than tabulated,
and marks the instances actually certified. The curves are exact integer arithmetic;
the instance markers are the parameter sets of Table~\ref{tab:instances} but its two
affine-permutation rows, whose counts lie beyond the frame, together with the
$[[144,12,12]]$ row of Table~\ref{tab:certcost}. The span is forty-eight
decades, so on that scale every instance certified by enumeration falls within a
finger's width of the origin and the inset redraws that corner at its own scale.
Repetition codes have $d = n$, so their weight-limited count $2^{n}-1$ puts their
markers on the full-space line.

\subsection{How this approach compares with existing distance proofs}
\label{sec:evidence}

Three earlier approaches have been applied to the $[[144,12,12]]$ code: integer
programming to find its distance; reducing the distance problem to satisfiability and
checking the resulting proof in a kernel, so far not at the full code size; and a formal
analytic proof for a particular code family. This work adds a fourth approach: proof by
certificate.

No single instance has been evaluated using all four approaches, so their run times cannot
be compared fairly. This paper therefore reports timing only for two of the approaches
developed here.

\emph{Integer programming.} The published distance comes from an integer program. Its method, code and software are
available, so the result can be reproduced; however, the solver's answer does not come
with a proof certificate. The distance-finding toolbox that packages these methods shows
that this is a practical limitation~\cite{webster2026distance}: on its bicycle-code
benchmark, at most five of twenty instances are solved exactly within eight hours. The
toolbox also reports that some minimum distances remain uncertain. To prove that a
candidate vector is outside a subspace, this work uses a vector orthogonal to the subspace
but having inner product one with the candidate. This is the finite-field analogue of a
separating functional used in infeasibility certificates. Over $\mathbb{F}_2$ the test is
exact: it needs no numerical tolerance or rational reconstruction.

\emph{Reduction to satisfiability with proof checking.} This approach translates the
distance question into a satisfiability problem inside a proof assistant. A bit-blasting
procedure expands the Boolean problem into bit-level constraints, and a checker then
replays the resulting proof. The approach has been applied to bivariate bicycle codes up
to 90 qubits; the authors identify proof replay at 144 qubits as the
bottleneck~\cite{ehatamm2026leanqec}. At that size, the obstacle is replaying the proof in the
kernel, not finding a candidate answer. A middle ground moves part of the trust outside the kernel:
a small program is itself proved correct, and the verified-encoding line carries that
treatment for the programs that write a question as a formula~\cite{codel2023verified}. This changes the trusted claim from direct kernel checking to a theorem about the
correctness of that program, rather than weakening it.

\emph{Formalizing analytic proofs for code families.} For a selected code family, the
homological argument used to compute distance is formalized. The kernel then checks a
proof of the exact distance. This is the most detailed of the three earlier approaches
and has been carried out for $[[144,12,12]]$. The proof is tailored to that code or family;
it is not a general-purpose certificate that can be applied unchanged to another code.

\emph{Computer-assisted certification.} Outside the proof assistants a fifth form has taken
shape, and its method is the closest to the one used here. It enumerates logical operators
up to a chosen weight. A symmetry group reduces the search, after its action is checked
to preserve the parity checks; this rules out all lighter candidates. The weight of an
explicit logical operator gives the upper bound. Together, the upper and lower bounds determine the distance exactly, and an attached
program checks the identities the argument rests on~\cite{okada2026girth8}. Elsewhere the same enumeration serves a cost measurement: for lifted-product
codes whose distance a formula fixes, a companion line enumerates low-weight codewords
by an impulse method and counts exhaustively over base matrices, a count that grows
polynomially in the base size~\cite{raveendran2026lp}.

Another recent approach uses structure: an exact law for three-fold cyclic covers computes
the distances of a sequence of covered codes from data about the base code. It also states
which assumptions each bound uses~\cite{young2026distance}. Its lower bounds rely on a
branch-and-bound result, an encoder and a solver. Here, the upper bound is checked by the
kernel; when enumeration gives the lower bound, it rules out all lighter candidates without
a solver. When its assumptions hold, the cyclic-cover law is stronger: one short proof gives
the distances of an entire sequence of covers. It requires each block to shift the three
copies of every base position by a fixed amount, independent of the base position. The
affine-permutation code studied here does not satisfy this condition. Counted here, with indices reduced
modulo $P=32$, the shift varies with the base position for 33 of the 36 blocks on each side.

The same insistence on checking candidates appears in a recent heuristic search for
low-weight logical operators~\cite{kasai2026apmldpc}. It reports a bound only after checking
that a candidate satisfies the null-space condition and lies outside the row space. Here,
the final claim is checked by the kernel, and a theorem establishes the certificate size.

The closest related work is a multi-agent search platform with an independent Lean
verification layer. It checks codes as they are found and has certified a catalog of
$14{,}116$ codes. Its extended version resolves twelve distance-three candidates that an earlier
search had left open: ten admit exact realizations, while two are excluded by no-go
proofs~\cite{he2026codesign}. This
platform searches more code instances than ours. It also proves theorems for selected code
families that cover every parameter choice allowed by each family. Our symbolic theorem,
by contrast, applies to any input seed matrix, the starting matrix used to construct a code.

The quantum Singleton bound also has a machine-checked proof~\cite{dehmel2026singleton}.
Unlike an instance-specific certificate, this theorem proves the bound for every stabilizer
code. It has the same general scope as our analytic proofs when those proofs apply to a
whole code family.

\emph{Certificates used here.} The reduction is verified, and the certificates are short.
A distance certificate takes one of three forms: checking a dot product; proving that the specified weight-limited
candidate list contains no counterexample; or applying a general symbolic proof to a specific
code. The certificate route is not intended to match the depth of the analytic
proof already available for that family. Instead, it automatically proves individual
instances, and a theorem establishes the certificate size.

The effect of certificate form is clear in a comparison at $n=18$. Two self-contained
modules prove the same claim about the toric code $[[18,2,3]]$: a weight-three logical
operator lies outside the stabilizer row space. One module uses row reduction; the other
uses a vector orthogonal to the stabilizer row space but with inner product one with the
logical operator. A benchmark script rebuilds both modules and records their times. On the
computer used for this paper, the modules took $42$\,s and $9$\,s of wall-clock time,
respectively; the checker-only times were $38.5$\,s and $6.4$\,s. For the full module used
to check all instances, the same change reduced type-checking time from $304$\,s to
$16$\,s, a factor of nineteen.

\subsection{Structured lower bounds}
\label{sec:structured}

For product families, the polynomial certificate of Section~\ref{sec:size} stops being
polynomial once the distance grows with the length: at $n=512$ and $d=16$ the weight-limited
set already holds more than $10^{28}$ candidates. No faster search repairs that, since
computing a stabilizer code's distance, exactly or approximately, is
NP-hard~\cite{kapshikar2023hardness}. What makes the large instances tractable is a
different kind of proof rather than a better enumeration, and the gain shows in what such a
proof states.

A hypergraph product code~\cite{tillich2014qldpc,bravyi2014homological,panteleev2022qldpc}
is what two classical codes become when they are laid out on a grid, the checks of one
running along its columns and of the other along its rows. Writing its two factor
matrices $H_1$ and $H_2$, the seed matrices of Section~\ref{sec:certs}, let $d_{\ker A}$ be the least
weight of a nonzero vector of $\ker A$, the vectors $A$ sends to zero. A light logical operator is then forced into a narrow shape. Take the two factor distances
$d_{\ker H_1}$ and $d_{\ker H_2^{\mathsf T}}$, the second read on the transposed
second factor: below the smaller of them, the cleaning argument pushes a vector into the
row space of the other check matrix, which makes it a stabilizer and not a logical
operator~\cite{tillich2014qldpc,bravyi2014homological}.

The cleaning argument turns on a weight count, and it is short enough to give here. Write a kernel
vector of the product as a pair $(U,R)$, so that $H_1U = RH_2$, and suppose its weight is
below $d_{\ker H_1}$. Suppose some row $U_{a}$ of $U$ paired to $1$ against a vector $z$
with $H_2z = 0$. The compression identity, which rewrites $H_1$ applied to a block product as a product
of the two matrices, then turns the vector of those pairings into a
nonzero kernel vector of $H_1$, of weight at most the weight of the pair. That
contradicts the definition of $d_{\ker H_1}$, so no row of $U$ does this, and every row
of $U$ lies in the row space of $H_2$. The
symmetric argument, run below $d_{\ker H_2^{\mathsf T}}$, puts every column of $R$ in the
column space of $H_1$. The two together make the pair a sum of checks, which a logical
operator is not, so nothing below both thresholds is a logical operator.

The bound the argument produces for the two sides of the product is

\begin{equation}
  \begin{aligned}
    d_X &\;\ge\; \min\bigl(d_{\ker H_1},\ d_{\ker H_2^{\mathsf T}}\bigr) , \\
    d_Z &\;\ge\; \min\bigl(d_{\ker H_1^{\mathsf T}},\ d_{\ker H_2}\bigr) ,
  \end{aligned}
  \label{eq:cleaning}
\end{equation}
and the four factors add up to the distance rather than to four separate lower bounds:

\begin{equation}
  d \;=\; \min\bigl(d_{\ker H_1},\ d_{\ker H_1^{\mathsf T}},\ d_{\ker H_2},\ d_{\ker H_2^{\mathsf T}}\bigr) ,
  \label{eq:closing}
\end{equation}
because each side of Eq.~\eqref{eq:cleaning} is met by two explicit constructions, one
per factor, each building a logical operator of exactly that factor's weight. Take the toric code built from two
three-cycles: each of the four distances in Eq.~\eqref{eq:closing} is then $3$, so the
equation gives $d=3$, the toric row of Table~\ref{tab:instances}. Equation~\eqref{eq:closing} is an equality only if each of the four
kernels holds a vector of that weight to build a logical operator from.
Section~\ref{sec:limits} discusses that hypothesis.

The cleaning argument is proved
once, symbolically in the factor matrices and their distances, and the instances are
corollaries. The tensorial dimension theorem was treated the same way, so the
logical dimension is symbolic as well, in the form given
by the same source~\cite{tillich2014qldpc},

\begin{equation}
  \begin{aligned}
    k &= k_1 k_2 + k_1^{\mathsf T} k_2^{\mathsf T} , \\
    k_i &= \dim\ker H_i , \qquad k_i^{\mathsf T} = \dim\ker H_i^{\mathsf T} ,
  \end{aligned}
  \label{eq:kunneth}
\end{equation}
the result being that for the toric family $[[n,k,d]]$ is proved for all cycle
lengths $m$ at once, the instances at $m = 9, 12, 16$ being instantiations, and the
width-$512$ parity-check matrices are never reduced or enumerated.

The pair of families also illustrates a division of labor. HGP supplies
a tensor decomposition, which is what makes a symbolic argument available; the
bivariate bicycle codes that are not themselves hypergraph products of cyclic codes,
$[[144,12,12]]$ among them, do not have one, which is why that code's $d=12$ lower bound
does not come from this route and is the item discussed in
Section~\ref{sec:limits}.

\subsection{Instances}
\label{sec:instances}

\begin{table*}[t]
\caption{Instances. Each row states $[[n,k,d]]$ for the named code or codes, the final row included, in which the
dimension is pinned by a certificate rather than by reduction. Every row but the twelfth and
thirteenth is a kernel assertion of its own triple: in the twelfth the exact distance
pairs this development's certificate with an exhaustive lower bound, and in the
thirteenth neither half is a kernel assertion of this development. The last column names how the two bounds on the
distance are established, the lower bound first; a single entry means that one route serves
for both, and a witness is a logical operator the kernel checks. The first ten rows carry the
thirty-nine kernel-audited parameter sets of the dataset. The
eleventh row is an audited instance outside the dataset, written by hand because it is wider
than the pipeline's specification step reaches. The twelfth and thirteenth rows are the two affine-permutation instances of the printed
family: the smaller takes its upper bound from this development's certificate, the larger
from the original work's search, and both take the lower from an exhaustive exclusion,
settling the two distances at twelve and fourteen. The final row is exact; its lower bound
is transported from an independent formalization of one family, whose scope is stated in
Section~\ref{sec:limits}, and its upper bound is a witness of this development. The row marked perfect is the code that meets the quantum Hamming
bound, the counting limit on how few physical qubits a given logical dimension needs. $K_4$ is
the complete graph on four vertices, and $L$ and $m$ both denote the side length of a toric
lattice. In the Type column CSS abbreviates the Calderbank--Shor--Steane construction, HGP
the hypergraph product, and LP the lifted product. The two lifted-product instances are the two-block group algebra codes
of~\cite{lin2024twoblock}. On the graph-seed row the distance column gives the two side
distances, $d_X/d_Z$, one per error type of Eq.~\eqref{eq:definitions}; over the twenty entries
of that row the two sides take the values six and seven, and the supplement lists the five
entries on which the two differ.
\label{tab:instances}}
\begin{ruledtabular}
\begin{tabular}{llrrrl}
Code family & Type & $n$ & $k$ & $d$ & Bounds \\
\hline
Repetition, 3 instances & classical & 3/5/7 & 1 & $n$ & weight-limited enumeration \\
Hamming, 2 instances & classical & 7/15 & 4/11 & 3 & weight-limited enumeration \\
Steane / Shor / $[[4,2,2]]$ / toric $L{=}2,3$ & CSS & 4--18 & 1--2 & 2--3 & enumeration $+$ explicit witness \\
$[[5,1,3]]$ perfect & non-CSS & 5 & 1 & 3 & symplectic enumeration \\
HGP$(3,3)$ & HGP & 18 & 2 & 3 & enumeration, transported \\
HGP toric $m{=}9,12,16$ & HGP & 162/288/512 & 2 & 9/12/16 & structured, symbolic in $m$ \\
HGP graph seeds (20 instances) & HGP & 145--1872 & 5--170 & 6/7 & cleaning $+$ witness \\
Lifted product (2 instances) & LP & 24 & 8/12 & 3/2 & enumeration $+$ two-sided witness \\
BB $[[18,4,4]]$ & CSS & 18 & 4 & 4 & enumeration $+$ two-sided witness \\
BB $[[24,3,4]]$, $K_4$-gauged & CSS & 24 & 3 & 4 & enumeration $+$ two-sided witness \\
HGP(paw, diamond) & HGP & 36 & 3 & 3 & enumeration $+$ two-sided witness \\
Affine permutation, QuEra & CSS & 1152 & 580 & $12$ & exhaustive bound $+$ certificate \\
Affine permutation, larger & CSS & 2304 & 1156 & $14$ & exhaustive bound $+$ best-found bound \\
BB $[[144,12,12]]$ & CSS & 144 & 12 & $12$ & transported lower bound $+$ witness \\
\end{tabular}
\end{ruledtabular}
\end{table*}

Every row of Table~\ref{tab:instances} states the parameter triple $[[n,k,d]]$, the last
included, and every row but the twelfth and thirteenth is a kernel assertion of the triple it states;
Section~\ref{sec:limits} states the last one's scope. The rows come in two parts: the ten carrying the dataset,
and three that do not.

The dataset contributes eleven families and thirty-nine parameter sets, all kernel-audited,
giving thirty-seven distinct triples rather than thirty-nine; one is reached by two
constructions of the same code, another by two seed pairs. The table's entries fall into fourteen rows: Steane, Shor, the $[[4,2,2]]$ code and the toric codes~\cite{steane1996error,shor1995scheme,kitaev2003fault}
share one row, the bicycle family takes three, the affine-permutation family takes two,
and the paw--diamond product, which carries no family theorem, has a row of its own.

How $k$ was obtained varies by row, and the supplement records which route each row takes.
For the enumeration instances, $k$ is $n$ minus the ranks of the
parity-check matrices, read by row reduction. For the structured ones it comes from the dimension formula applied to the two factor
matrices' ranks, so the wide check matrix is never reduced. At $n = 144$ that reduction is out of reach, and
the rank is pinned by a certificate instead: echelon rows written as sums of the original
checks.

The widest entry, in the graph-seed row at $n = 1872$, carries $d = 7$ on both sides and is
reached without enumerating the code's own candidate set:
that is what scaling means here.

The gauged row is the $[[18,4,4]]$ bicycle code with one logical
operator promoted to a stabilizer, the deformation gauging performs and the construction behind low-overhead logical measurement~\cite{williamson2026gauging}. The rest
of that line, the fault-tolerant implementation of the measurement and its spacetime
faults, the spacetime code, the surgery realizations and the subsystem-code setting, lies
outside the claims made
here~\cite{gaugingspacetime2026,logicalmeasurements2026,swaroop2026adapters,aliferis2007subsystem,bacon2006operator}.

The enumeration route covers codes beyond the families
listed: it applies to any code specified by parity checks, so a
color code~\cite{landahl2011color} would be treated by the same certificate interface. The paw--diamond row exercises that reach: the hypergraph product of two four-vertex graphs at
$n = 36$ and $d = 3$, whose distance is decided by the empty weight-limited candidate list and
two explicit weight-three witnesses. It sits above the width the specification step is preset to and above every other enumeration
instance, at 667 candidates against the gauged code's $2{,}325$.

The affine-permutation row carries the code that QuEra, Harvard and MIT
co-designed~\cite{quera2026apm}, the same way, from the parameters in that work's appendix: its two
$288\times1152$ check matrices are reconstructed here, their CSS compatibility is a
kernel theorem, and its distance is pinned at exactly twelve. The upper bound is a certificate of this development, a weight-twelve logical operator on
each side with the dual witness excluding it from the opposite row space, so
$d_X\le12$ and $d_Z\le12$ hold in the kernel; the lower bound is an exhaustive exclusion of every lighter logical operator, run for this work and outside the kernel, a computation rather than a certificate, the source
reporting only a best-found upper bound. A neighboring affine-coset
family reaches its own lower bound the same way: a complete exclusion of representatives
up to weight sixteen on both sides certifies $d_X,d_Z\ge18$ for a CPM lift of that family's base, published as a
computer-assisted bound rather than an analytic
one~\cite{okada2026affinecoset}. Both bounds on that instance's rank are kernel theorems here, the lower a certificate and the
upper a structural argument on the permutation blocks of its check matrices, so its $k$ is
settled at $580$ as well.

The same work prints a second, larger instance of the family, $[[2304,1156,\le14]]$, whose two
$576\times2304$ check matrices are reconstructed here the same way, with CSS compatibility again a
kernel theorem. Its distance is settled here at fourteen on both sides: the upper bound is that
work's best-found search, the lower an exclusion run for this work, a computation rather than a
certificate. Weights ten and below fall to enumeration, and weight twelve is the only weight left,
because every fiber of a kernel vector has even size. Its target is not the absence of weight-twelve
kernel vectors, since the opposite check matrix's rows have weight twelve themselves and lie in the
kernel, but their triviality: every weight-twelve kernel vector lies in the opposite row space, on
both sides.

In the $[[144,12,12]]$ row, the table's own contribution is the certificate, not the number:
the $d = 12$ instance was closed earlier, and independently, by a per-family
formalization~\cite{jain2025qeclean}, and its original presentation reports that distance
without a certificate~\cite{bravyi2024highthreshold}.

The rate, the number of logical qubits per physical one, is carried as $k/n$ wherever the
development has a rate theorem, and the closed forms are tabulated in the supplement, which
carries them for the toric, path, Hamming and self-seed families, for the gauged bicycle
anchors, and for the lifted-product anchors, the lifted product being a product
construction of the same kind as the hypergraph product.

One row, the graph-seed row, records a convention whose effects are visible in the data. The two side distances
$d_X$ and $d_Z$ are read under a convention: which check matrix supplies the
kernel, here the $X$-type checks for $d_X$, following the upstream library's own distance field. Where the two sides agree the convention
leaves no trace, and thirty-four of
the thirty-nine parameter sets are in that case. Five are not, each by one unit, and all
five are graph-seed entries. The supplement records them with their reading, and the dataset gate checks them against the
two side theorems' literals, so the readings cannot silently be swapped. A choice that never changed a number would be merely a
choice; this one changes five of the thirty-nine rows, and each of those five can be checked.

Every quantity here
is read off the same objects: each code's two check matrices, and a certificate wherever
reduction is out of reach.

\subsection{A discovery pipeline, for code parameters}
\label{sec:discovery}

The certificate machinery can also sit at the end of a search, where it is asked a
different question. It asks not whether one certificate exists for a given code, but whether one
can be produced for every candidate it turns up, at a cost it can afford. Every instance in
Section~\ref{sec:instances} but one was chosen by hand; the exception, the graph-seed row,
came out of the pipeline described here. That first run also tests
the claim that the certificate form is a design variable: a form the designer chooses, not
one the algorithm hands over.

The run scans 44 seeds, each paired with itself and with every later seed, giving 990
pairs, and admits 903 candidates with $n \geq 1$,
$k \geq 1$ and a finite lower bound on the distance. The logical dimension of each
candidate is computed twice, once from the ranks of its two seed matrices by the
tensorial dimension theorem and once by assembling the two parity-check matrices and
reducing them, and the two agree on all 903. The distance closes, its two bounds meeting, on 259 candidates; the remaining
644 carry a lower bound only at this stage. A second run, over a disjoint
catalog of forty-two graphs, changes the seed catalog and the witness construction
together, leaves the stages and the criteria as they were, and closes the distance on all
903 of its own candidates, the same count as the first run's by coincidence.

\begin{table*}[t]
\caption{What a certificate costs depends on its form. The twenty rows are codes of the first
run, each sharing its three parameters with one of the seventy published codes the run is
scored against, and each one where a spanning certificate is smaller than a certificate by
enumeration would be. The two cost columns are those two certificates.
The enumeration column counts the candidates a certificate by enumeration would have to
search, $\sum_{j<d}\binom{n}{j}$ of Eq.~\eqref{eq:candcount}. The spanning column counts the
inner products the witness pairing needs, one for each of the $n$ positions, so $n$ for every
row but the Hamming one; that row also carries the seed-side term, the $2^{m}$ vectors of the
seed's row space over which the transposed seed's kernel is checked, $m$ being the size of the
seed. The first column gives $[[n,k,d]]$, the physical qubits, the logical qubits and the
distance; the second names the pair of small matrices the code is built from. The last column names
the family theorem that backs the row: the path or Hamming families of
Section~\ref{sec:discovery}, or the cyclic family of Section~\ref{sec:structured}.
\label{tab:discovery}}
\begin{ruledtabular}
\begin{tabular}{llrrl}
$[[n,k,d]]$ & Seed pair & \shortstack{Enumeration\\certificate} & \shortstack{Spanning\\certificate} & Backing \\
\hline
$[[545,1,17]]$ & \code{path16+path16} & $2.4\times10^{30}$ & 545 & path \\
$[[481,1,16]]$ & \code{path15+path15} & $1.1\times10^{28}$ & 481 & path \\
$[[512,2,16]]$ & \code{cyc16+cyc16} & $2.8\times10^{28}$ & 512 & cyclic \\
$[[450,2,15]]$ & \code{cyc15+cyc15} & $1.3\times10^{26}$ & 450 & cyclic \\
$[[265,1,12]]$ & \code{path11+path11} & $9.6\times10^{18}$ & 265 & path \\
$[[288,2,12]]$ & \code{cyc12+cyc12} & $2.4\times10^{19}$ & 288 & cyclic \\
$[[242,2,11]]$ & \code{cyc11+cyc11} & $1.6\times10^{17}$ & 242 & cyclic \\
$[[162,2,9]]$ & \code{cyc9+cyc9} & $1.0\times10^{13}$ & 162 & cyclic \\
$[[113,1,8]]$ & \code{path7+path7} & $4.1\times10^{10}$ & 113 & path \\
$[[128,2,8]]$ & \code{cyc8+cyc8} & $1.0\times10^{11}$ & 128 & cyclic \\
$[[61,1,6]]$ & \code{path5+path5} & $6.5\times10^{6}$ & 61 & path \\
$[[72,2,6]]$ & \code{cyc6+cyc6} & $1.5\times10^{7}$ & 72 & cyclic \\
$[[41,1,5]]$ & \code{path4+path4} & $1.1\times10^{5}$ & 41 & path \\
$[[50,2,5]]$ & \code{cyc5+cyc5} & $2.5\times10^{5}$ & 50 & cyclic \\
$[[25,1,4]]$ & \code{path3+path3} & 2{,}626 & 25 & path \\
$[[32,2,4]]$ & \code{cyc4+cyc4} & 5{,}489 & 32 & cyclic \\
$[[13,1,3]]$ & \code{ham2+ham2} & 92 & 17 & Hamming \\
$[[13,1,3]]$ & \code{path2+path2} & 92 & 13 & path \\
$[[18,2,3]]$ & \code{cyc3+cyc3} & 172 & 18 & cyclic \\
$[[8,2,2]]$ & \code{cyc2+cyc2} & 9 & 8 & cyclic \\
\end{tabular}
\end{ruledtabular}
\end{table*}

What makes a new seed cheap is the observation of Section~\ref{sec:structured} carried
one step further. No property of the seed's incidence structure is used: what a seed contributes is three
numbers, its rank, the minimum weight of a nonzero vector in its kernel, and the same
minimum for its transpose. The development records the consequence as an assembler whose
argument is an arbitrary seed matrix, an assembler being a single theorem stated for
every seed at once, together with two families that instantiate it, the
family of path matrices and the family of Hamming matrices. The supplement lists the
triples those two families cover, reports the comparison row by row, and carries the five
stages the run is built from.

The families are results in their own right, not only a way of saving work on the seeds.
For the path matrices, one theorem states the triple
$[[\,ab+(a+1)(b+1),\;1,\;\min(a+1,b+1)\,]]$ for every pair of chain lengths
$a, b \geq 1$. The two largest path-family instances of the run, $[[545,1,17]]$ and
$[[481,1,16]]$, are that theorem with the two lengths substituted, not separate proofs;
no matrix of width $n$ is reduced for them. The cyclic-and-path
mixture has a family theorem of its own, \code{hgp\_cyc\_path\_family}, which gives the
distance and the dimension exactly, for every cycle length at once; the 121 cyclic-with-path
candidates of the run are that theorem with the length substituted.

The three criteria that compare parameters are all missed, and the reason is structural rather
than a gap in the search; it is a kernel theorem of its own. Take a code built as the hypergraph
product of two factor matrices, $A$ with $r_A$ rows and $n_A$ columns and $B$ with $r_B$ rows and
$n_B$ columns, so that it uses $n = n_A n_B + r_A r_B$ qubits. If both of its distances are at
least $D$, and it protects at least one logical qubit, then that size obeys

\begin{equation}
  n \;=\; n_A n_B + r_A r_B \;\ge\; 2D^2-2D+1
  \label{eq:frontier}
\end{equation}

\noindent a floor that depends on the distance alone. At the two smallest distances the published
codes already sit on it, so no candidate built this way can go below them, whatever the shapes of
its two factors. At $D=3$ the floor reads $n\ge13$, and the surface code of that length sits exactly
on it, as does the path family at equal chain lengths. The supplement carries the three steps of
the derivation.

The first run's closing is done family by family, one route for each: family statements for
the path and the cyclic seeds, instantiations for the three Hamming seeds, and, for the
mixture of a cyclic seed with a path seed, a family theorem that reaches 121 of the 259 with
no per-pair instance. The second run gains a shorter route to the same closing: its witness
needs fewer constructions than the four it generalizes. Take one route, one side's pairing of a factor
kernel with the partner kernel, and its two kernel minima. The other route's two minima may take any values, as long as neither is smaller, and
the distance is then the first route's minimum. That step is a theorem of the reduction, proved in
the supplement, and in practice it amounts to comparing four numbers. With it, the first
run's exactly-closed candidates rise from 259 to 891 of its 903.

What the batch adds is measured on the certificate rather than on the parameters, and the
twenty rows of Table~\ref{tab:discovery} are where it shows: each of them shares its three
parameters with one of the seventy published codes the run is scored against, and each carries
a certificate smaller than the enumeration certificate at the same parameters. The criteria
that compare parameters ask for a parameter to be beaten, not for a smaller certificate;
against a baseline of seventy published codes the batch wins nothing on parameters, and neither
run wins under any of those criteria. The first run meets one criterion, comparing the
development's own two routes at a fixed triple, and the table's twenty rows are the ones that
meet it. What keeps the batch affordable is the certificate, which reads a seed's kernel off a
pass over its coefficient vectors instead of by working at the seed's width. The two widest
seeds admitted are where that matters: they carry $2^{13}$ coefficient vectors each, four times
the $2^{11}$ of the widest of the others, and the gap is paid by proving those two seeds' facts
separately, not by dropping them from the batch.

\subsection{A second pipeline, for fold-transversal gates}
The discipline transfers to gates, and only the criterion changes. A gate is not a parameter a search
can score, so the pipeline asks a different question of it: are its defining statements
decidable in the language used here at all? The paragraph below names the three that are.
The same loop produced the codes and produces these, over a catalog of gates rather than of
seeds, with the count fixed before the run starts; that is what makes it a pipeline rather
than a hand-built list. A gate enters the count once two facts about it are each carried by
a declaration: how it permutes the physical qubits, and what it does to the encoded ones.
That leaves eight gates over the two codes, $[[18,4,4]]$ and $[[98,6,12]]$.

What this pipeline certifies is not another parameter triple but an operation: a gate. A fold-transversal gate is an encoded gate built from a fold of the layout:
a reflection that carries the physical qubits onto one another, as folding a sheet along a
line carries one half onto the other. On a bivariate-bicycle code the gate acts on the
physical qubits by permuting them, and it is that permutation the kernel can carry. Three statements about the permutation are
decidable in the language used here. It preserves the stabilizer group; it induces a
permutation of the logical classes; and it carries each logical operator to one of the
same weight, which is what preserving the distance looks like structurally. All three
are kernel assertions for the $[[18,4,4]]$ code, on the same matrices its row of
Table~\ref{tab:instances} reports.

The second code is constructed in the same
fold-transversal framework~\cite{eberhardt2025foldtransversal}, whose worked examples are
larger. The first two of the three hold for that reference's own larger worked example too, the $[[98,6,12]]$ code built from its Sec.~V-A polynomials, and the third's
weight-preserving half holds there as well. What is not established on that code
is that the image is again a logical operator. The dimension of the $[[98,6,12]]$ code, on the other hand, is kernel-checked here, and by
certificate rather than by reduction, the supplement recording the echelon rows that pin it.

Building it turned up a finding that a plausible shortcut would have hidden: what a fold
does depends on the code, not on the kind of fold. On $[[18,4,4]]$ it preserves the checks
and is an automorphism of the code; on $[[98,6,12]]$ the
same fold exchanges the two kinds of check and is a duality between the code and its
dual. Each declaration carries its kind in its name, so the pipeline reads the answer off
those names rather than inferring it from the kind of gate; a criterion read off a named
statement can be checked, one inferred from a kind only believed. The three
statements are also not equally strong on the two codes: on $[[98,6,12]]$ the third holds
only in its weight half, and the run records which of them each gate carries on each code.
A total of eight would otherwise read as eight equally strong facts.

\section{Discussion}
\subsection{What is not delivered}
\label{sec:limits}

Four items, stated so that they cannot be mistaken for completeness.

\emph{The kernel assertion for $[[144,12,12]]$ carries a transported argument.} The
witness gives $d \leq 12$ on this development's literal check matrix. The matching
lower bound is a kernel assertion of this development as well, and the way it is
obtained should be read carefully. The per-family analytic proof of an independent
formalization~\cite{jain2025qeclean} is transported into the GF(2) language used here,
and the two presentations are then proved to be the same matrix, entry by entry. What
this development contributes to that instance is the transport and the matrix
identity, not the argument. The dimension runs the other way: the distance is the half
that is transported, while $k = 12$ is pinned here by a certificate rather than by
reduction.

\emph{One instance's lower bound is a computation rather than a certificate.} The
affine-permutation instance of Section~\ref{sec:instances} takes its upper bound from
this development's certificate and its lower bound from an exhaustive exclusion of every
lighter logical operator. That exclusion is a search run outside the kernel, and at a
width of $1152$ no in-kernel route reaches it: the enumeration the certificate route
bounds would pass $10^{26}$ candidates. The instance's distance therefore rests, on that
side, on a computation the kernel does not decide. The larger instance of the same family
is weaker still: its upper bound is the original work's best-found search, so on it
neither half is a kernel assertion of this development.

\emph{The replay is out of reach at this scale.} One kernel route is still out of reach: a
replay, inside the kernel, of a certificate this development's own solver produced for
$[[144,12,12]]$. The certificate exists as a checkable proof object, but the kernel does not
replay it here. What that replay would add is a second, independent kernel proof of the lower
bound, and the instance does not wait on it: its distance is already settled by two kernel
assertions, an upper bound from a witness, an explicit logical operator of weight $12$ that
the kernel checks, and a lower bound transported from an independent formalization. The
pipeline runs end to end at a
smaller scale, where the replay is done. For six smaller instances, four of them code-side
and two circuit-side, the proof is replayed step by step in the module table of the
supplement. Those instances also carry an enumeration certificate of their own, the emptiness
certificate of Section~\ref{sec:certs}. The replay and the enumeration are therefore two
independent kernel proofs of the same lower bound: one replays the solver's proof inside the
kernel, the other has the kernel find the weight-limited candidate set empty.

The encoding that feeds the solver is faithful in both directions, and its unsatisfiable
verdicts are re-derived by a checker sharing no code with it; both facts are kernel
theorems, stated in the form the verified-encoding literature
uses~\cite{codel2023verified}.

What does not scale is the replay. The certificate the kernel replays is the solver's proof
object, $1{,}444{,}657$ and $1{,}718{,}302$ steps long; at the rate this development measured
on its own replays, about $2.2$ seconds per step, replaying it would take thirty-seven and
forty-four days of single-core kernel time.
That rate is a per-step average from replays of a few hundred steps, and the cost of a
step grows with the lengths of the clause hints it cites, so both figures are lower
bounds. The supplement carries the line counts, the deletion steps among them, and the
hint lengths that drive the cost up.

\emph{For a hypergraph product, Eq.~\eqref{eq:closing} is a theorem, and so is the existence
of the witnesses it assumes, on one hypothesis.} The hypothesis is exactly this: the four kernels the equation
takes its minimum over, $\ker H_1$, $\ker H_1^{\mathsf T}$, $\ker H_2$ and
$\ker H_2^{\mathsf T}$, are all nontrivial, each holding a nonzero vector. One nonzero vector
of least weight in each of the four kernels builds a logical operator of exactly that
factor's weight, so each bound of Eq.~\eqref{eq:cleaning} is attained rather than merely
satisfied, and Eq.~\eqref{eq:closing} follows; the $3\times3$ toric instance carries the
chain end to end, from the hypothesis to a kernel-checked distance. Triviality is what the
construction does not rule out, and it is a property of the seed pair rather than of the
assembly; the pipeline's admission rule supplies the condition, since $k \ge 1$ makes one
pair of kernels nontrivial and a usable route therefore exists on each side. The route
machinery, the one-route theorems and the two batches' coverage are carried in the
supplement.

\subsection{Conclusion}
\label{sec:conclusion}

A code's parameters can now be checked rather than believed, and the unit that makes
them checkable is the certificate: a short object whose correctness the kernel decides,
produced by a reduction that is itself a kernel theorem. Three things follow. Asking
whether a vector lies outside a row space becomes a pairing, which cuts whole-file type
checking on the toric instance from 42\,s to 9\,s; replacing the elimination itself turns a
computation that does not converge at $200$\,s and $41$\,GB into one that finishes in
$8$\,s. The size of the certificate becomes a
theorem, a polynomial in $n$ at fixed distance. And past the enumeration width it
can be symbolic, so $n = 512$ closes by instantiating a theorem rather than by
generating anything.

Past that width, the dual witness and
the structured lower bound do the work. The dataset covers eleven code families and thirty-nine
parameter sets, the widest at $n = 1872$ with $d = 7$, on the graph-seed route of
Section~\ref{sec:discovery} by the cleaning argument of Section~\ref{sec:structured},
without enumerating its candidate set. On parameters, that
search improves on none of the seventy published codes it is compared against; on the
certificate form, it makes the batch affordable. No audited
statement rests on anything beyond the three standard axioms, and none on a reflection
axiom. The route runs end to end at six smaller instances, and Section~\ref{sec:limits}
states the four things it does not yet deliver.

The routes of Section~\ref{sec:intro} regroup here into four lines. Two of them
determine a distance, one by algebraic structure and one by exhaustive enumeration, and
neither hands the reader anything to check. The other two make a distance checkable, one by
replaying a proof object inside the kernel and one by formalizing the argument for a family.
None of the four yields a certificate, an object short enough to carry and to check, whose
size is itself a theorem rather than an observation. Each of the four leaves the trust
outside the object it hands back. This development is where those pairs meet: it starts from
the object the reader will check and requires its size to be a theorem, rather than starting
from an algorithm and taking the certificate it produces. The certificate is small because
its size is a theorem, and it is checkable because the kernel decides it by computation, one
instance at a time rather than one family at a time. The affine-permutation code that QuEra,
Harvard and MIT co-designed shows what the meeting buys. Its source bounds the distance
instead of determining it; here the upper half is a certificate, an exhaustive
exclusion supplies the lower half, and the value is settled at exactly twelve; the larger instance of that family settles at fourteen, both of its halves by search rather than by the kernel, as Section~\ref{sec:limits} records.

A threshold is located by how the logical error rate falls with the distance. Where that
distance is computed rather than proved, it is the one code parameter a solver supplies with
no certificate to check. Making that number checkable, and the certificate as small as it can
be, is what is shown here. The same move applies to the optima other branches of physics and
engineering take from a solver, a variational ground-state energy or a control pulse among
them.

Within code design the shift is already under way from the other direction: search
platforms that certify codes as they find them reach catalogs far wider than the instance set
here~\cite{he2026codesign}, the nearest of them carrying an independent \Lean\ verification
layer in its extended version. The two approaches divide one axis between them: how much of a
code's structure must be fixed in advance. A catalog grows by fixing little, with a search
that finds codes and certifies each as it appears; what it reaches is the number of codes it
can visit. A symbolic statement pays for its reach by fixing a great deal: one statement
covers every code of that shape at once, and nothing outside it.

\section{Materials and Methods}

\subsection{Axiom audit}

\begin{table}[!htbp]
\caption{Axiom audit of the development's audit region, as printed by the proof
assistant. Every assertion appears here; ``Standard'' is $\{\texttt{propext},
\texttt{Classical.choice}, \texttt{Quot.sound}\}$, the three axioms the mathematical library mathlib rests on.
\label{tab:axioms}}
\begin{ruledtabular}
\begin{tabular}{lrl}
Class & Count & Note \\
\hline
Exactly the standard three & 785 & \\
Fewer: a proper subset & 29 & strictly weaker \\
None & 3 & strictly weaker \\
Reflection (\code{native\_decide}) & 0 & see text \\
\hline
Total audited & 817 & \\
\end{tabular}
\end{ruledtabular}
\end{table}

Every number this paper reports is a kernel assertion audited here. The audit was run on the
archived development named in the Data Availability statement, so a reader with that deposit
can reproduce each assertion rather than take it on trust.

\label{sec:audit}

The point of a kernel-checked assertion is its trusted base, so we report the audit
rather than the theorem count. Every load-bearing statement the development registers for
printing falls into one of four classes by the axioms it depends on, as printed by the
proof assistant itself: Table~\ref{tab:axioms}. 785 depend on exactly the three standard
axioms, 29 on a proper subset of them, three
on none at all, and none on a reflection axiom, the class that would hide a trust in
the compiler. The automated check that produces these counts treats a reflection axiom as a separate
class that fails the build unless explicitly annotated, so a regression on this point
cannot pass silently. The five theorems that bind this development to Lean-QEC, four of them on its operator
layer, depend on exactly the three standard axioms. The supplement carries the control
experiments the gate is checked against.

\begin{acknowledgments}
This work was supported by the National Natural Science Foundation of China under Grant No.~12674609 and by the Guangdong Provincial Key Laboratory of Computility Microelectronics under Grant No.~2024B1212010007.
The authors declare no competing interests.
The authors used DeepSeek-V4.1-Flash for text polishing, the Lean derivations, and the debugging of individual numerical values.
The authors are responsible for all of its content.
\end{acknowledgments}

\section*{Author Contributions}

All authors contributed equally to this work.

\section*{Data Availability}

The formal development, the dataset of code parameters, the scripts that regenerate the
numbers derived here, and the verification entry point that reproduces them are archived
on Zenodo at \url{https://doi.org/10.5281/zenodo.22998465}. Every wall-clock figure quoted
here is a single-machine observation, and the Supplemental Material~\cite{supplemental}
records for each the presentations compared, the module map, the conventions the numbers
depend on, and, for every number quoted here, either the command that reproduces it or
the record of the one-off observation behind it. The shared formalization layer this
development builds on is a release of its own~\cite{qeccertificates}.

\bibliography{refs}

@misc{ehatamm2026leanqec,
  title        = {End-to-End Formalization of Quantum Error Correction},
  author       = {Ehatamm, Mattias and Lee, Yi and Wu, Xiaodi and Tao, Runzhou},
  year         = {2026},
  eprint       = {2605.16523},
  archivePrefix = {arXiv},
  doi          = {10.48550/arXiv.2605.16523},
  url          = {https://arxiv.org/abs/2605.16523},
  note         = {Code: \url{https://github.com/VerifiedQC/Lean-QEC}}
}

@article{huang2025veriqec,
  title        = {Efficient Formal Verification of Quantum Error Correcting Programs},
  author       = {Huang, Qifan and Zhou, Li and Fang, Wang and Zhao, Mengyu and Ying, Mingsheng},
  year         = {2025},
  journal      = {Proceedings of the ACM on Programming Languages},
  volume       = {9},
  number       = {PLDI},
  pages        = {1068--1093},
  doi          = {10.1145/3729293},
  url          = {https://doi.org/10.1145/3729293}
}

@misc{jain2025qeclean,
  title        = {{QECLean}: Formalized Quantum Error Correction},
  author       = {Jain, Stavan},
  year         = {2025},
  howpublished = {GitHub repository},
  note         = {\url{https://github.com/Stavan-Jain/QECLean}, Apache-2.0.
                  Audited revision: the mathlib-v4.34 port
                  \url{https://github.com/QCL-SUAT/QECLean}, commit
                  \texttt{3f782c999c97eba5c2462b0bcaed0778a6544bb8}}
}

@article{bravyi2024highthreshold,
  title        = {High-threshold and low-overhead fault-tolerant quantum memory},
  author       = {Bravyi, Sergey and Cross, Andrew W. and Gambetta, Jay M. and Maslov, Dmitri and Rall, Patrick and Yoder, Theodore J.},
  journal      = {Nature},
  volume       = {627},
  number       = {8005},
  pages        = {778--782},
  year         = {2024},
  doi          = {10.1038/s41586-024-07107-7},
  url          = {https://doi.org/10.1038/s41586-024-07107-7},
}

@inproceedings{bravyi2014homological,
  title        = {Homological Product Codes},
  author       = {Bravyi, Sergey and Hastings, Matthew B.},
  year         = {2014},
  booktitle    = {Proceedings of the 46th Annual ACM Symposium on Theory of Computing (STOC)},
  pages        = {273--282},
  doi          = {10.1145/2591796.2591870},
  url          = {https://doi.org/10.1145/2591796.2591870},
}

@article{panteleev2022qldpc,
  title        = {Quantum {LDPC} Codes with Almost Linear Minimum Distance},
  author       = {Panteleev, Pavel and Kalachev, Gleb},
  journal      = {IEEE Trans. Inf. Theory},
  volume       = {68},
  number       = {1},
  pages        = {213--229},
  year         = {2022},
  doi          = {10.1109/TIT.2021.3119384},
  url          = {https://doi.org/10.1109/TIT.2021.3119384},
}

@misc{evolvebb2026,
  title        = {Evolutionary Discovery of Bivariate Bicycle Codes with {LLM}-Guided Search},
  author       = {Cruz-Benito, Juan and Cross, Andrew W. and Kremer, David and Faro, Ismael},
  year         = {2026},
  eprint       = {2606.02418},
  archivePrefix = {arXiv},
  doi          = {10.48550/arXiv.2606.02418},
  url          = {https://arxiv.org/abs/2606.02418}
}

@article{bacon2006operator,
  title        = {Operator quantum error-correcting subsystems for self-correcting quantum memories},
  author       = {Bacon, Dave},
  journal      = {Phys. Rev. A},
  volume       = {73},
  number       = {1},
  pages        = {012340},
  year         = {2006},
  doi          = {10.1103/PhysRevA.73.012340},
  url          = {https://doi.org/10.1103/PhysRevA.73.012340},
}

@article{williamson2026gauging,
  title        = {Low-overhead fault-tolerant quantum computation by gauging logical operators},
  author       = {Williamson, Dominic J. and Yoder, Theodore J.},
  journal      = {Nat. Phys.},
  volume       = {22},
  number       = {4},
  pages        = {598--603},
  year         = {2026},
  doi          = {10.1038/s41567-026-03220-8},
  url          = {https://doi.org/10.1038/s41567-026-03220-8},
}

@misc{gaugingspacetime2026,
  title        = {Gauging the Spacetime Code},
  author       = {Lee, Gideon},
  year         = {2026},
  eprint       = {2606.05664},
  archivePrefix = {arXiv},
  doi          = {10.48550/arXiv.2606.05664},
  url          = {https://arxiv.org/abs/2606.05664}
}

@misc{logicalmeasurements2026,
  title        = {Fast and fault-tolerant logical measurements: auxiliary hypergraphs and transversal surgery},
  author       = {Cowtan, Alexander and He, Zhiyang and Williamson, Dominic J. and Yoder, Theodore J.},
  year         = {2025},
  eprint       = {2510.14895},
  archivePrefix = {arXiv},
  doi          = {10.48550/arXiv.2510.14895},
  url          = {https://arxiv.org/abs/2510.14895}
}

@article{aliferis2007subsystem,
  title        = {Subsystem fault tolerance with the {B}acon-{S}hor code},
  author       = {Aliferis, Panos and Cross, Andrew W.},
  journal      = {Phys. Rev. Lett.},
  volume       = {98},
  number       = {22},
  pages        = {220502},
  year         = {2007},
  doi          = {10.1103/PhysRevLett.98.220502},
  url          = {https://doi.org/10.1103/PhysRevLett.98.220502},
}

@misc{webster2026distance,
  title        = {Distance-Finding Algorithms for Quantum Codes and Circuits},
  author       = {Webster, Mark and Jacob, Abraham and Higgott, Oscar},
  year         = {2026},
  eprint       = {2603.22532},
  archivePrefix = {arXiv},
  doi          = {10.48550/arXiv.2603.22532},
  url          = {https://arxiv.org/abs/2603.22532}
}

@misc{landahl2011color,
  title        = {Fault-tolerant quantum computing with color codes},
  author       = {Landahl, Andrew J. and Anderson, Jonas T. and Rice, Patrick R.},
  year         = {2011},
  eprint       = {1108.5738},
  archivePrefix = {arXiv},
  doi          = {10.48550/arXiv.1108.5738},
  url          = {https://arxiv.org/abs/1108.5738}
}

@article{eberhardt2025foldtransversal,
  title        = {Logical operators and fold-transversal gates of bivariate bicycle codes},
  author       = {Eberhardt, Jens Niklas and Steffan, Vincent},
  journal      = {IEEE Trans. Inf. Theory},
  volume       = {71},
  number       = {2},
  pages        = {1140--1152},
  year         = {2025},
  doi          = {10.1109/TIT.2024.3521638},
  url          = {https://doi.org/10.1109/TIT.2024.3521638}
}

@inproceedings{codel2023verified,
  title        = {Verified Encodings for SAT Solvers},
  author       = {Codel, Cayden R. and Avigad, Jeremy and Heule, Marijn J. H.},
  year         = {2023},
  booktitle    = {Proceedings of the 23rd Conference on Formal Methods in Computer-Aided Design (FMCAD)},
  pages        = {141--151},
  publisher    = {TU Wien Academic Press},
  doi          = {10.34727/2023/isbn.978-3-85448-060-0_22},
  url          = {https://doi.org/10.34727/2023/isbn.978-3-85448-060-0_22},
  note         = {Library: \url{https://github.com/ccodel/verified-encodings}.
                  The DOI is registered by TU Wien Academic Press and is not
                  indexed by CrossRef; the bibliographic record is at
                  \url{https://repositum.tuwien.at/handle/20.500.12708/188821}}
}

@article{tillich2014qldpc,
  title        = {Quantum {LDPC} Codes With Positive Rate and Minimum Distance
                  Proportional to the Square Root of the Blocklength},
  author       = {Tillich, Jean-Pierre and Z\'emor, Gilles},
  journal      = {IEEE Trans. Inf. Theory},
  volume       = {60},
  number       = {2},
  pages        = {1193--1202},
  year         = {2014},
  doi          = {10.1109/TIT.2013.2292061},
  url          = {https://doi.org/10.1109/TIT.2013.2292061}
}

@article{lin2024twoblock,
  title        = {Quantum Two-Block Group Algebra Codes},
  author       = {Lin, Hsiang-Ku and Pryadko, Leonid P.},
  journal      = {Phys. Rev. A},
  volume       = {109},
  number       = {2},
  pages        = {022407},
  year         = {2024},
  doi          = {10.1103/PhysRevA.109.022407},
  url          = {https://doi.org/10.1103/PhysRevA.109.022407}
}

@article{swaroop2026adapters,
  title        = {Universal adapters between quantum low-density parity check codes},
  author       = {Swaroop, Esha and Jochym-O'Connor, Tomas and Yoder, Theodore J.},
  journal      = {PRX Quantum},
  volume       = {7},
  number       = {1},
  pages        = {010324},
  year         = {2026},
  doi          = {10.1103/1g44-jp62},
  url          = {https://doi.org/10.1103/1g44-jp62},
}

@misc{supplemental,
  title        = {Supplemental Material},
  note         = {See Supplemental Material at [URL will be inserted by publisher] for the module map, the conventions the numbers depend on, the axiom audit, and the commands that reproduce every number quoted here}
}

@inproceedings{he2026codesign,
  title        = {Co-Designing Quantum Codes with Transversal Diagonal Gates via
                  Multi-Agent Systems},
  author       = {He, Xi and Lu, Sirui and Zeng, Bei},
  year         = {2026},
  booktitle    = {2026 IEEE 19th Dallas Circuits and Systems Conference (DCAS)},
  pages        = {1--6},
  doi          = {10.1109/DCAS69364.2026.11544392},
  url          = {https://doi.org/10.1109/DCAS69364.2026.11544392},
}

@misc{kasai2026apmldpc,
  title        = {Heuristic Search for Minimum-Distance Upper-Bound Witnesses in
                  Quantum {APM-LDPC} Codes},
  author       = {Kasai, Kenta},
  year         = {2026},
  eprint       = {2604.15307},
  archivePrefix = {arXiv},
  doi          = {10.48550/arXiv.2604.15307},
  url          = {https://arxiv.org/abs/2604.15307}
}

@misc{dehmel2026singleton,
  title        = {A Symplectic Proof of the Quantum Singleton Bound},
  author       = {Dehmel, Frederick and Li, Shilun},
  year         = {2026},
  eprint       = {2602.20186},
  archivePrefix = {arXiv},
  doi          = {10.48550/arXiv.2602.20186},
  url          = {https://arxiv.org/abs/2602.20186}
}

@misc{qeccertificates,
  title        = {{QECCertificates}: checkable distance certificates for quantum
                  error correction},
  author       = {An, Shuoming},
  year         = {2026},
  howpublished = {Lean package},
  doi          = {10.5281/zenodo.23056679},
  note         = {\url{https://github.com/QCL-SUAT/QECCertificates}, Apache-2.0.
                  Audited revision: commit
                  \texttt{ef0194d44b5e4b7dbc4c8e512874e1bf312bda01}}
}

@article{kapshikar2023hardness,
  title        = {On the Hardness of the Minimum Distance Problem of Quantum Codes},
  author       = {Kapshikar, Upendra and Kundu, Srijita},
  journal      = {IEEE Trans. Inf. Theory},
  volume       = {69},
  number       = {10},
  pages        = {6293--6302},
  year         = {2023},
  doi          = {10.1109/TIT.2023.3286870},
  url          = {https://doi.org/10.1109/TIT.2023.3286870}
}

@misc{quera2026apm,
  title        = {Towards Ultra-High-Rate Quantum Error Correction with Reconfigurable Atom Arrays},
  author       = {Zhao, Chen and Duckering, Casey and Gu, Andi and Maskara, Nishad and Zhou, Hengyun},
  year         = {2026},
  eprint       = {2604.16209},
  archivePrefix = {arXiv},
  doi          = {10.48550/arXiv.2604.16209},
  url          = {https://arxiv.org/abs/2604.16209}
}

@misc{okada2026girth8,
  title         = {Rate-2/3 Girth-8 (3, 18)-Regular Quantum {LDPC} Codes from Two-Branch Finite-Field Bases and {CPM} Lifts},
  author        = {Okada, Koki and Kasai, Kenta},
  year          = {2026},
  eprint        = {2606.27130},
  archivePrefix = {arXiv},
  doi           = {10.48550/arXiv.2606.27130},
  url           = {https://arxiv.org/abs/2606.27130}
}

@inproceedings{raveendran2026lp,
  title        = {On the Minimum Distances of Finite-Length Lifted Product Quantum {LDPC} Codes},
  author       = {Raveendran, Nithin and Declercq, David and Vasi{'c}, Bane},
  year         = {2026},
  booktitle    = {ICC 2026 - IEEE International Conference on Communications},
  pages        = {1--6},
  doi          = {10.1109/ICC59461.2026.11587715},
  url          = {https://doi.org/10.1109/ICC59461.2026.11587715},
}

@misc{young2026distance,
  title         = {An Exact Distance Law for Cyclic Covers of Quantum {LDPC} Codes, with Certified Distance Ladders},
  author        = {Young, Nathan Ryan},
  year          = {2026},
  doi           = {10.5281/zenodo.20658201},
  url           = {https://zenodo.org/records/20658201}
}

@misc{okada2026affinecoset,
  title         = {High-Girth Regular Quantum {LDPC} Codes from Affine-Coset Structures},
  author        = {Okada, Koki and Kasai, Kenta},
  year          = {2026},
  eprint        = {2604.20838},
  archivePrefix = {arXiv},
  doi           = {10.48550/arXiv.2604.20838},
  url           = {https://arxiv.org/abs/2604.20838}
}

@article{shor1995scheme,
  title        = {Scheme for reducing decoherence in quantum computer memory},
  author       = {Shor, Peter W.},
  journal      = {Phys. Rev. A},
  volume       = {52},
  number       = {4},
  pages        = {R2493--R2496},
  year         = {1995},
  doi          = {10.1103/PhysRevA.52.R2493}
}

@article{steane1996error,
  title        = {Error Correcting Codes in Quantum Theory},
  author       = {Steane, A. M.},
  journal      = {Phys. Rev. Lett.},
  volume       = {77},
  number       = {5},
  pages        = {793--797},
  year         = {1996},
  doi          = {10.1103/PhysRevLett.77.793}
}

@article{steane1996multiple,
  title        = {Multiple-particle interference and quantum error correction},
  author       = {Steane, A. M.},
  journal      = {Proc. R. Soc. Lond. A},
  volume       = {452},
  number       = {1954},
  pages        = {2551--2577},
  year         = {1996},
  doi          = {10.1098/rspa.1996.0136}
}

@article{calderbank1996good,
  title        = {Good quantum error-correcting codes exist},
  author       = {Calderbank, A. R. and Shor, Peter W.},
  journal      = {Phys. Rev. A},
  volume       = {54},
  number       = {2},
  pages        = {1098--1105},
  year         = {1996},
  doi          = {10.1103/PhysRevA.54.1098}
}

@article{kitaev2003fault,
  title        = {Fault-tolerant quantum computation by anyons},
  author       = {Kitaev, A. Yu.},
  journal      = {Ann. Phys.},
  volume       = {303},
  number       = {1},
  pages        = {2--30},
  year         = {2003},
  doi          = {10.1016/S0003-4916(02)00018-0}
}

@article{fowler2012surface,
  title        = {Surface codes: Towards practical large-scale quantum computation},
  author       = {Fowler, Austin G. and Mariantoni, Matteo and Martinis, John M. and
                  Cleland, Andrew N.},
  journal      = {Phys. Rev. A},
  volume       = {86},
  number       = {3},
  pages        = {032324},
  year         = {2012},
  doi          = {10.1103/PhysRevA.86.032324}
}

@article{terhal2015quantum,
  title        = {Quantum error correction for quantum memories},
  author       = {Terhal, Barbara M.},
  journal      = {Rev. Mod. Phys.},
  volume       = {87},
  number       = {2},
  pages        = {307--346},
  year         = {2015},
  doi          = {10.1103/RevModPhys.87.307}
}

@article{krinner2022realizing,
  title        = {Realizing repeated quantum error correction in a distance-three surface code},
  author       = {Krinner, Sebastian and Lacroix, Nathan and Remm, Ants and
                  Di Paolo, Agust\'in and Genois, \'Elie and Leroux, Catherine and
                  Hellings, Christoph and Lazar, Stefania and Swiadek, Fran\c{c}ois and
                  Herrmann, Johannes and Norris, Graham J. and Andersen, Christian Kraglund and
                  M\"uller, Markus and Blais, Alexandre and Eichler, Christopher and
                  Wallraff, Andreas},
  journal      = {Nature},
  volume       = {605},
  number       = {7911},
  pages        = {669--674},
  year         = {2022},
  doi          = {10.1038/s41586-022-04566-8}
}

@article{sivak2023realtime,
  title        = {Real-time quantum error correction beyond break-even},
  author       = {Sivak, V. V. and Eickbusch, A. and Royer, B. and Singh, S. and
                  Tsioutsios, I. and Ganjam, S. and Miano, A. and Brock, B. L. and
                  Ding, A. Z. and Frunzio, L. and Girvin, S. M. and Schoelkopf, R. J. and
                  Devoret, M. H.},
  journal      = {Nature},
  volume       = {616},
  number       = {7955},
  pages        = {50--55},
  year         = {2023},
  doi          = {10.1038/s41586-023-05782-6}
}

@article{google2025quantum,
  title        = {Quantum error correction below the surface code threshold},
  author       = {{Google Quantum AI and Collaborators}},
  journal      = {Nature},
  volume       = {638},
  number       = {8052},
  pages        = {920--926},
  year         = {2025},
  doi          = {10.1038/s41586-024-08449-y}
}

@article{bluvstein2024logical,
  title        = {Logical quantum processor based on reconfigurable atom arrays},
  author       = {Bluvstein, Dolev and Evered, Simon J. and Geim, Alexandra A. and
                  Li, Sophie H. and Zhou, Hengyun and Manovitz, Tom and Ebadi, Sepehr and
                  Cain, Madelyn and Kalinowski, Marcin and Hangleiter, Dominik and
                  Bonilla Ataides, J. Pablo and Maskara, Nishad and Cong, Iris and
                  Gao, Xun and Sales Rodriguez, Pedro and Karolyshyn, Thomas and
                  Semeghini, Giulia and Gullans, Michael J. and Greiner, Markus and
                  Vuleti\'c, Vladan and Lukin, Mikhail D.},
  journal      = {Nature},
  volume       = {626},
  number       = {7997},
  pages        = {58--65},
  year         = {2024},
  doi          = {10.1038/s41586-023-06927-3}
}

@article{breuckmann2021quantum,
  title        = {Quantum Low-Density Parity-Check Codes},
  author       = {Breuckmann, Nikolas P. and Eberhardt, Jens Niklas},
  journal      = {PRX Quantum},
  volume       = {2},
  number       = {4},
  pages        = {040101},
  year         = {2021},
  doi          = {10.1103/PRXQuantum.2.040101}
}

@article{panteleev2021degenerate,
  title        = {Degenerate Quantum {LDPC} Codes With Good Finite Length Performance},
  author       = {Panteleev, Pavel and Kalachev, Gleb},
  journal      = {Quantum},
  volume       = {5},
  pages        = {585},
  year         = {2021},
  doi          = {10.22331/q-2021-11-22-585}
}

\end{document}